\documentclass[twocolumn,secnumarabic,amssymb, aps,prb,reprint]{revtex4-1}
\usepackage{graphicx}
\usepackage{color}
\usepackage{dcolumn}
\usepackage{bm}
\usepackage{ucs}
\usepackage{physics}
\usepackage[modulo]{lineno}
\linenumbers
\usepackage{multirow}

\usepackage{mathtools}
\begin{document}
\nolinenumbers

\title{Doubling the mobility of InAs/InGaAs selective area grown nanowires}

\author{Daria V. Beznasyuk$^{1}$}
\email{daria.beznasiuk@nbi.ku.dk}
\author{Sara Mart\'{i}-S\'{a}nchez$^{2}$}
\author{Jung-Hyun Kang$^{1}$}
\author{Rawa Tanta$^{1}$}
\author{Mohana Rajpalke$^{3}$}
\author{Toma\v{s} Stankevi\v{c}$^{3}$}
\author{Anna Wulff Christensen$^{3}$}
\author{Maria Chiara Spadaro$^{2}$}
\author{Roberto Bergamaschini$^{4}$}
\author{Nikhil N. Maka$^{1}$}
\author{Christian Emanuel N. Petersen$^{5}$}
\author{Damon J. Carrad$^{1,5}$}
\author{Thomas Sand Jespersen$^{1,5}$}
\author{Jordi Arbiol$^{2,6}$}
\author{Peter Krogstrup$^{1,3}$}

\affiliation{$^{1}$Center for Quantum Devices, Niels Bohr Institute, University of Copenhagen, 2100 Copenhagen, Denmark}
\affiliation{$^{2}$Catalan Institute of Nanoscience and Nanotechnology (ICN2), CSIC and BIST, Campus UAB, Bellaterra, Barcelona, Catalonia, Spain}
\affiliation{$^{3}$Microsoft Quantum Materials Lab Copenhagen, 2800 Lyngby, Denmark}
\affiliation{$^{4}$L-NESS and Dipartimento di Scienza dei Materiali, Universit\`a di Milano-Bicocca, I-20125 Milano, Italy}
\affiliation{$^{5}$ICREA, Passeig de Lluís Companys 23, 08010 Barcelona, Catalonia, Spain}
\affiliation{$^{6}$Department of Energy Conversion and Storage, Technical University of Denmark, Fysikvej, Building 310, 2800 Lyngby}

\date{\today}

\begin{abstract}

Selective area growth (SAG) of nanowires and networks promise a route toward scalable electronics, photonics and quantum devices based on III-V semiconductor materials. The potential of high-mobility SAG nanowires however is not yet fully realized, since interfacial roughness, misfit dislocations at the nanowire/substrate interface and non-uniform composition due to material intermixing all scatter electrons. Here, we explore SAG of highly lattice-mismatched InAs nanowires on insulating GaAs(001) substrates and address these key challenges. Atomically smooth nanowire/substrate interfaces are achieved with the use of atomic hydrogen (a-H) as an alternative to conventional thermal annealing for the native oxide removal. The problem of high lattice mismatch is addressed through an In$_x$Ga$_{1-x}$As buffer layer introduced between the InAs transport channel and the GaAs substrate. The Ga-In material intermixing observed in both the buffer layer and the channel is inhibited via careful tuning of the growth temperature. Performing scanning transmission electron microscopy and x-ray diffraction analysis along with low-temperature transport measurements we show that optimized In-rich buffer layers promote high quality InAs transport channels with the field-effect electron mobility over~10000~cm$^2$V$^{-1}$s$^{-1}$. This is twice as high as for non-optimized samples and among the highest reported for InAs selective area grown nanostructures.

\end{abstract}

\maketitle
\section{Introduction}

For the past decades semiconductor nanostructures have been an important materials platform for nanoelectronics and mesoscopic quantum transport~\cite{Samuelson2003Oct,Zhang2015Oct,Lutchyn2018May}. The nanoscale confinement can be realized following various routes: by local electrostatic gating of two-dimensional heterostructures, using top-down processing, or by vapor-liquid-solid growth, where nanostructures grow out of plane of a substrate and the confinement is achieved by a nanoscale catalyst particle. Recently, selective area growth (SAG) of semiconductor structures and heterostructures has emerged as an appealing platform for the realization of electronic, optoelectronic and photonic devices~\cite{Pastorek2018Nov,Krizek2018Sep,Tutuncuoglu2015Nov,Conesa-Boj2012Dec,Park2018Oct,Yeh2012Jan}. In the SAG approach, the material growth occurs in lithographically predefined openings formed in a layer of amorphous mask on a semiconductor substrate. The advantages of SAG include control over shapes, dimensions, positions and faceting of the final structures~\cite{Wang2019Jun,Winnerl2019Mar}. Moreover, improved crystal quality of SAG materials as compared to their planar counterparts has been demonstrated~\cite{Hsu2012Dec,Park2018Oct}.

\begin{figure*} [t!]
\includegraphics[width=0.7\linewidth]{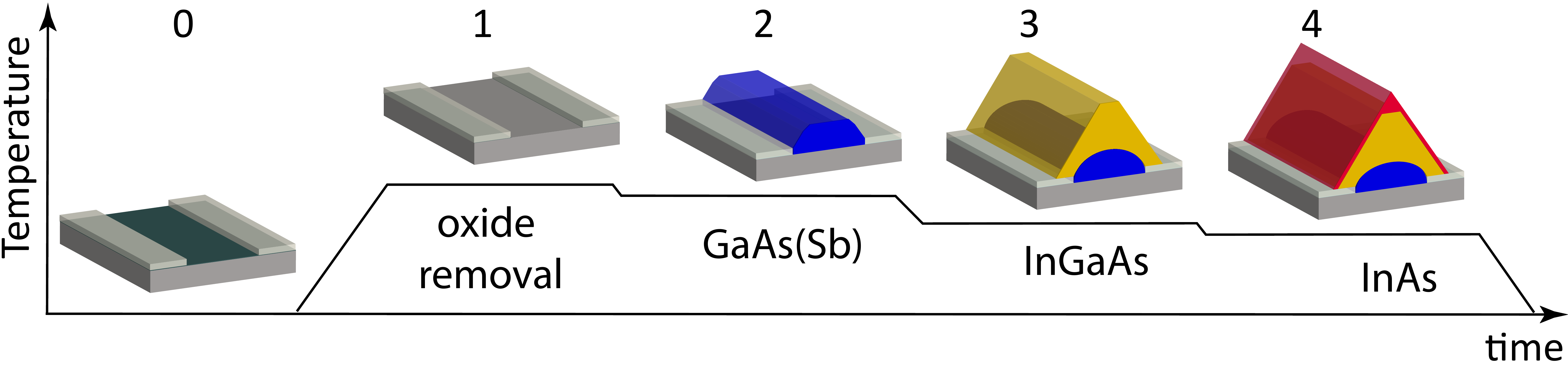}
\caption{Schematics of growth steps of InAs/InGaAs/GaAs(Sb) SAG nanowires on GaAs(001) substrates covered with a 10-nm thick SiO$_2$ mask. 0. Substrate fabrication; 1. Native oxide removal; 2. GaAs(Sb), 3. InGaAs and 4. InAs layer growth. Note that the originally faceted GaAs(Sb) buffer becomes rounded after the growth of InGaAs and InAs layers on top.}
\label{Fig1}
\end{figure*}

The SAG approach has been used to successfully grow out-of-plane nanowires and nanofins as well as in-plane nanowires, nanomembranes, nanoprisms, nanorings and quantum dots combining low-bandgap III-V and more exotic II-V materials with high-bandgap semiconductor substrates~\cite{Yeh2012Jan,Wang2019Jun, EscobarSteinvall2021Oct,Seidl2019Jul}. 
In-plane SAG of InAs and InSb nanowires attracts special attention for applications in quantum transport as controllable and scalable nanowire networks can be readily achieved~\cite{Fahed2016Nov,Gooth2017Apr,Desplanque2018May,Krizek2018Sep,Friedl2018Apr,Lee2019Jan,Aseev2019Dec,Aseev2019Jan,Liu2020Jan,OphetVeld2020Mar,Friedl2020May}. Since structures are already grown horizontally in plane of the substrate, it also simplifies their device processing.

However, while the growth of continuous SAG nanowires and networks hosting ballistic transport thought the junctions has recently been demonstrated~\cite{Gooth2017Apr}, issues related to surface/interface quality and material intermixing limit the electron mobility. 

For example, the substrate fabrication process combined with the native oxide removal in the mask windows by thermal annealing prior nanowire growth also provoke interfacial roughness and voids in the substrate~\cite{Krizek2018Sep,Tutuncuoglu2015Nov,Bucamp2019Apr}. Although surface relaxation allows for the growth of highly lattice-mismatched materials, networks of misfit dislocations may still occur for certain materials combinations, dimensions and growth conditions~\cite{Krizek2018Sep,Lee2019Aug, Aseev2019Jan}. A way to address this issue is to use buffer layers to accommodate the mismatch, as recently demonstrated for GaSb buffer layers between InAs SAG nanowires on GaAs substrates~\cite{Fahed2016Nov,Pastorek2018Nov}. However, the GaSb buffer is electrically conducting and it complicates applications in transport devices. Finally, material intermixing between layers is inherent to heteroepitaxial systems and it leads to degrading mobility~\cite{Basu1984Jul,Chin1991Oct,Gerard1992Mar,Muraki1993Feb,Joyce1998Dec,Liao1999Dec,Chaparro2000Jun}. Recent reports on SAG of InAs/GaAs nanowires showed dramatic 50-80\% Ga in nominally pure InAs active regions~\cite{Friedl2018Apr,Friedl2020May}.

In this work, we focus on SAG of highly lattice-mismatched InAs nanowires grown by means of molecular beam epitaxy (MBE) on GaAs(001) substrates and address the challenges stated above. We introduce an In$_x$Ga$_{1-x}$As buffer between conducting InAs and insulating GaAs and improve substrate cleaning prior to growth. Owing to material intermixing, In$_x$Ga$_{1-x}$As buffers are highly diluted with Ga, and nominally pure InAs channels consist of ternary In(Ga)As alloys prior to the optimization. By reducing the growth temperature, we suppress the Ga-In material intermixing achieving In$_x$Ga$_{1-x}$As buffer layers with high In content and pure InAs channels. In-rich buffers promote InAs/In$_x$Ga$_{1-x}$As interfaces with high crystal quality. This is directly reflected through measured electron mobility, which is doubled as compared to the non-optimized samples, reaching record high values of~12550~cm$^2$V$^{-1}$s$^{-1}$ for selective area grown InAs nanostructures~\cite{Tomioka2015Nov,Krizek2018Sep,Seidl2019Jul}. 

\section{Results and Discussion}

The sequence of growth steps leading to high quality InAs SAG nanowires is shown in Fig.~\ref{Fig1}. We use undoped GaAs(001) substrates. The nanowire geometry is controlled by defining windows in a 10 nm silicon dioxide (SiO$_2$) mask layer. The substrate fabrication (step 0) follows the Ref.~\cite{Krizek2018Sep} except for the etching process. We use inductively coupled plasma with a mixture of tetrafluoromethane (CF$_4$) and hydrogen (H$_2$) gases to reveal the pattern instead of hydrofluoric acid for better controllability of the process. The growth details for each step are discussed in Supplemental Material S1. 




\begin{figure*}
  \includegraphics[width=1\linewidth]{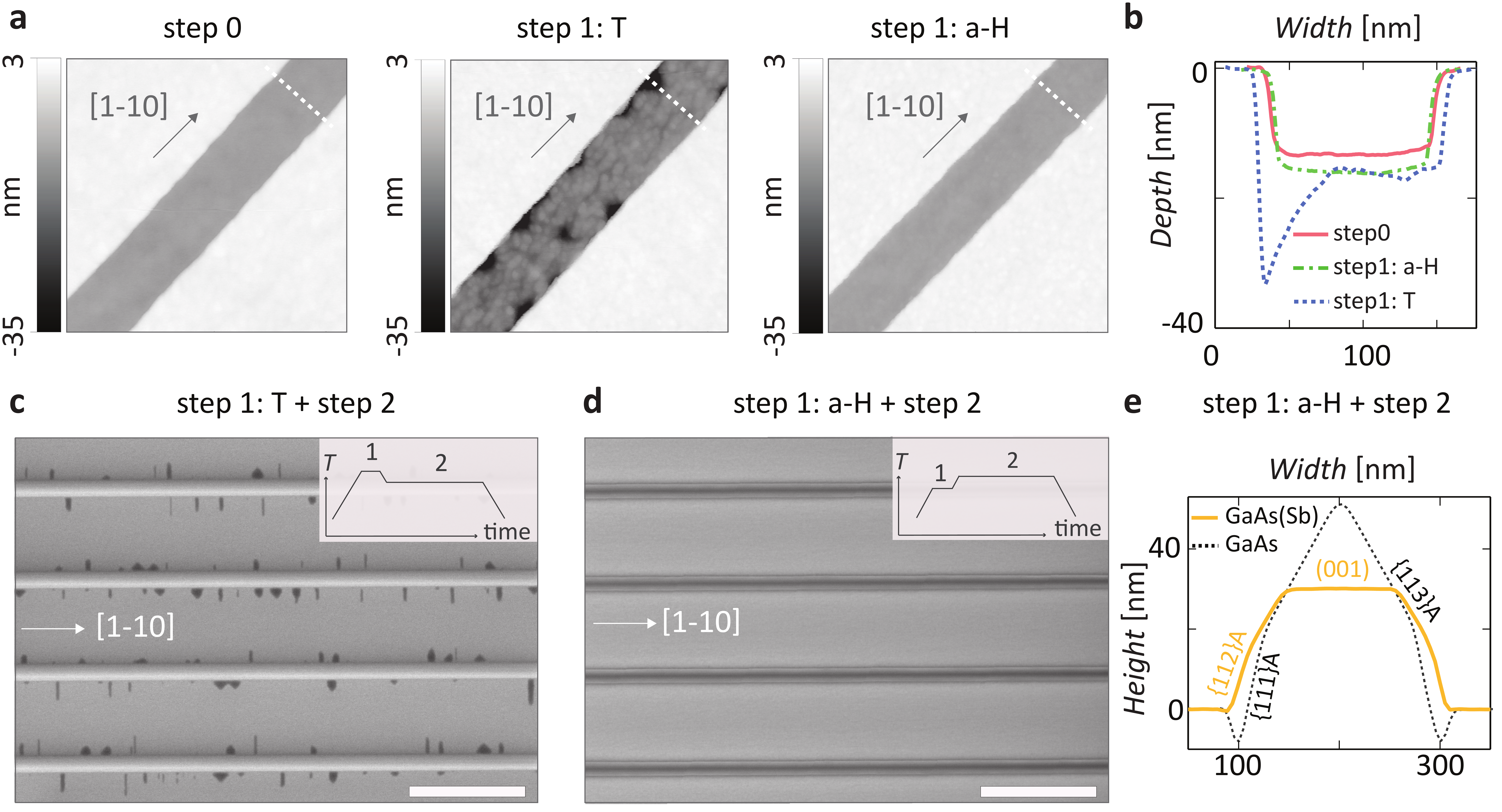}
  \caption{(a) A set of AFM images (0.5$\times$0.5 $\mu$m$^2$ image size) demonstrating a [1$\bar{1}$0]-oriented GaAs growth window (trench) with a native oxide (step 0), after thermal oxide desorption (step 1: T), after atomic hydrogen oxide removal (step 1: a-H). White dashed lines indicate the position of AFM profiles shown in (b) and averaged over 15 nm. Top-view SEM images (1 $\mu$m scale bar) of GaAs(Sb) SAG nanowires grown on a GaAs(001) substrate from which the native oxide was removed either thermally (c) or with a-H (d). The insets show relative temperature T of oxide removal with respect to the GaAs(Sb) growth step with the latter being fixed at~600~$^{\circ}$C. (e) AFM profile across a typical GaAs(Sb) SAG nanowire with step 1: a-H showing a faceted structure. AFM profile across a GaAs SAG nanowire grown without Sb surfactant but otherwise identical growth parameters is shown for comparison.}
  \label{Fig2}
\end{figure*} 

\subsection{Oxide removal and GaAs(Sb) buffer growth}
We first started by optimizing the procedure of removing the native oxide from the bottom of GaAs windows prior nanowire growth (step 1, Fig.~\ref{Fig1}). The standard approach for the oxide removal is thermal annealing which degrades the surface due to the temperature activated transformation of the stable Ga$_2$O$_3$ into the volatile Ga$_2$O by consumption of GaAs~\cite{SpringThorpe1987Jan, VanBuuren1991Jul, Smith1991Dec, Tone1992Jun, Adamcyk2000Jun}. The results are 15-30 nm deep pits as seen in Fig.~\ref{Fig2}a,b (``step 1:T") and Supplemental Material~S2. The root-mean-square (rms) roughness of the GaAs surface in the growth windows increases from 0.31$\pm$0.11 nm after etching to 3.18$\pm$0.35 nm after annealing. In conventional thin film epitaxy, the surface topography after thermal annealing can be improved by growing thick buffer layers. In the SAG samples, however, we find that the pits underneath the oxide mask often do not get filled during growth leaving voids easily distinguished in scanning electron microscopy (SEM) (black stripes in Fig.~\ref{Fig2}c)~\cite{Tutuncuoglu2015Nov, Krizek2018Sep, Aseev2019Jan, Friedl2020May}. These voids can potentially degrade SAG-based device performances and thus are unacceptable for many device applications.

There are several alternative ways of removing the native oxide from GaAs including group-III assisted annealing~\cite{Asaoka2003Apr,Li2011Aug} or atomic hydrogen (a-H)~\cite{Yamada1993Jun}. Here, we investigated the use of a-H which is known to produce atomically smooth GaAs surfaces in thin film epitaxy owing to significantly reduced temperatures required for the Ga$_2$O$_3$ to Ga$_2$O transformation and at the same time it also facilitates the reduction of the carbon contamination~\cite{Kawabe1995May,Bucamp2019Apr}. As shown in Fig.~\ref{Fig2}a,b (``step 1:a-H"), exposing GaAs growth windows to~3.0$\times$10$^{-5}$ mbar a-H at~350~$^{\circ}$C for~15~min results in atomically smooth GaAs surfaces with~0.37$\pm$0.14 nm rms roughness without affecting the surrounding oxide mask. Our findings show that a-H is an ideal approach to smoothly remove the native oxide from substrates partially covered by an oxide mask used in SAG.

The second step in the process is the growth of a GaAs(Sb) buffer layer where Sb is used as surfactant. It was found that this step is essential for improving electronic transport properties of InAs/GaAs SAG nanowire devices~\cite{Krizek2018Sep}. GaAs(Sb) nanowires grown on substrates prepared with a-H are shown in Fig.~\ref{Fig2}d. The nanowires grow continuously indicating that the native oxide was removed successfully. Figure~\ref{Fig2}e shows the topography across a typical GaAs(Sb) nanowire grown in a 220 nm wide [1$\bar{1}$0]-oriented growth window measured with Atomic Force Microscopy (AFM). The nanowire has a height of 30 nm above the trench and exhibits a flat top (001) facet with the rms roughness of 0.275$\pm$0.015 nm. The side facets were assigned to \{113\}A and \{112\}A crystal planes based on the slope (Supplemental Material~S3). We note that \{112\}A facets are not expected from the equilibrium crystal shape (ECS) model for GaAs grown without any surfactants~\cite{Lee2008Jan,Yeu2019Feb,Tobias2020Jul}. Instead, low-energy \{111\}A facets are predicted and observed experimentally (Fig.~\ref{Fig2}e)~\cite{Sato2004Aug}. One possible explanation is that the facet angle is underestimated in our experiment owing to only a small fraction of the side facet being available. On the other hand, it is likely that the presence of Sb surfactant changes both thermodynamics (reducing the surface energy) and kinetics increasing the surface diffusion of Ga adatoms on the facets) of the growing nanowires~\cite{Portavoce2004Apr,Nimmatoori2008Dec,Anyebe2015Apr,Yuan2015Sep,Borg2013Apr,Ren2016Feb}. 

\subsection{In$_x$Ga$_{1-x}$As buffer growth}

\begin{figure}
\centering
  \includegraphics[width=1\linewidth]{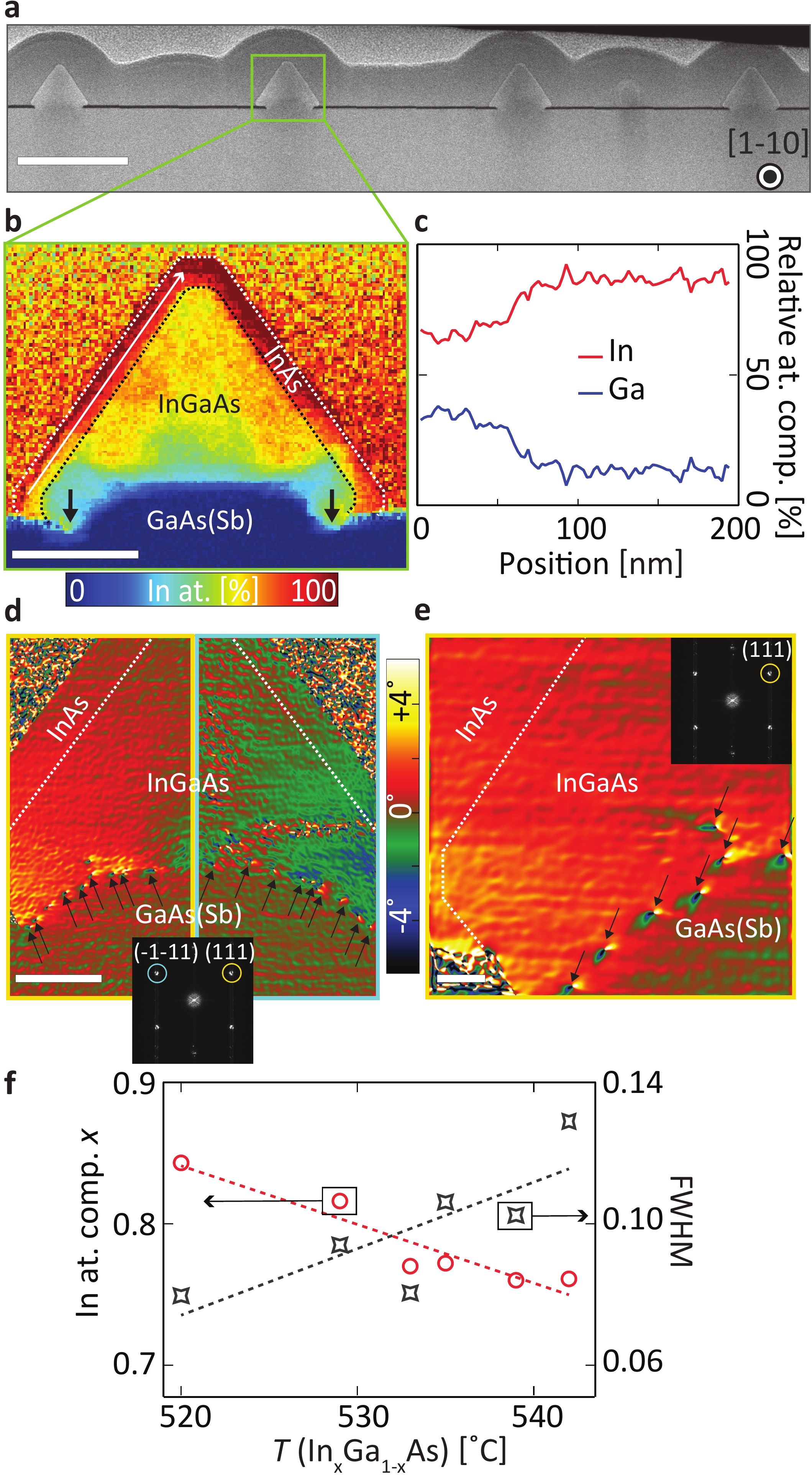}
  \caption{The role of InGaAs growth temperature on the composition and crystal properties of the InAs/InGaAs/GaAs(Sb) SAG nanowires. (a) Low magnification HAADF-STEM image (scale bar 500~nm) of a lamella taken across [1$\bar{1}$0]-oriented field-effect nanowires grown at~522~$^{\circ}$C. (b) In atomic distribution EELS map (relative to Ga, in atomic percentage, scale bar 100~nm) of the nanowire from (a). Arrows indicate the place of erosion of GaAs(Sb) and GaAs. (c) Relative Ga and In atomic composition profile along the ($\bar{1}\bar{1}$1) facet in the InAs channel from (b). (d),(e) GPA rotational maps of the nanowire from (b) representing its central part (50~nm scale bar) and the left corner (10~nm scale bar). Misfit dislocations at the InGaAs/GaAs(Sb) buffers are highlighted with black arrows. The insets show the corresponding fast Fourier transform (FFT) with the analysed planes: (111) planes correspond to the left half and ($\bar{1}\bar{1}$1) planes to the right half of the image in (d). (f) \emph{x} in the buffer as a function of the buffer growth temperature, T, extracted from XRD reciprocal space maps. The peak broadening representative of the compositional variations in the buffer measured as FWHM is plotted as well. The dashed lines are linear fits and are guides to the eye.}
  \label{Fig3}
\end{figure}

To accommodate the lattice mismatch between the GaAs substrate and the InAs active region and to trap misfit dislocations away from InAs, we introduce an In$_x$Ga$_{1-x}$As buffer layer sandwiched between GaAs and InAs (step 3, Fig.~\ref{Fig1}). Such approach is extensively used in planar semiconductor structures~\cite{Grider1998Jun,Inoue1991May}, but it has not yet been adopted for SAG nanowires. A high In content, \emph{x}, is required for the lattice matching. At the same time, the buffer must be electrically insulating which favours lower \emph{x}, leading to the need for a compromise. A set of InAs/In$_x$Ga$_{1-x}$As/GaAs(Sb) samples were grown in a temperature range from~520~$^{\circ}$C to~540~$^{\circ}$C to investigate the Ga-In material intermixing and in the following \emph{x}=0.9 was kept fixed. The bounds of the temperature range were dictated by the selectivity window~\cite{Aseev2019Jan}. The growth parameters of GaAs(Sb) and InAs layers were kept identical for all the samples (Supplemental Material~S1).

Figure~\ref{Fig3}a shows an STEM micrograph using the high-angle annular dark-field imaging mode (HAADF) taken across four [1$\bar{1}$0]-oriented InAs/InGaAs/GaAs(Sb) nanowires grown at~522~$^{\circ}$C. The nanowires protrude out of the growth window with a slight lateral overgrowth on the SiO$_2$ mask (the layer with the darkest contrast in this imaging mode). The cross section is symmetrically formed by the (001) top facet, \{111\}A (and small inclusions of \{111\}B), \{113\}A and \{110\} side facets as can be seen in STEM images and supported by simulated atomic model in Supplemental Material~S4~\cite{Momma2011Dec} and the ECS model~\cite{Tobias2020Jul}. Note that the InGaAs layer is grown without any surfactants and the \{111\} family of facets is observed in place of the \{112\} family seen for the first buffer layer. The EELS analysis in Fig.~\ref{Fig3}b reveals the presence of the GaAs(Sb) buffer layer, followed by an In$_x$Ga$_{1-x}$As region and a $\sim$15-20-nm thick InAs layer. Generally, the GaAs(Sb) layer has a rounded shape without the clear side facets observed earlier in the reference GaAs(Sb) nanowires (Fig.~\ref{Fig2}e). Moreover, its surface in the vicinity of the substrate as well as the substrate are eroded (black arrows in Fig.~\ref{Fig3}b). The In$_x$Ga$_{1-x}$As buffer layer has two regions with a distinct composition: the bottom Ga-rich part with x$\sim$0.5 at the InGaAs/GaAs(Sb) interface (visible as cyan flames in Fig.~\ref{Fig3}b) and the upper In-rich part with  x$\sim$0.8. Additional EELS maps are shown in Supplemental Material~S4. 
Finally, Fig.~\ref{Fig3}c shows the In composition profile extracted along the ($\bar{1}\bar{1}$1)A nanowire facet (white arrow in Fig.~\ref{Fig3}b) inside the nominally pure InAs channel. \emph{x} drops from $\sim$0.9 at the top of the nanowire to x$\sim$0.65 on its sides at the interface with the Ga-rich buffer region. This suggests that there is a constant flow of Ga which diffuses towards areas with abrupt change in composition, and thus highly strained, creating diluted InGaAs alloys. As shown later, the InAs/InGaAs interface is mainly dislocation free suggesting that the intermixing provides efficient strain relief for the system. 

\begin{figure*}
\includegraphics[width=1.0\linewidth]{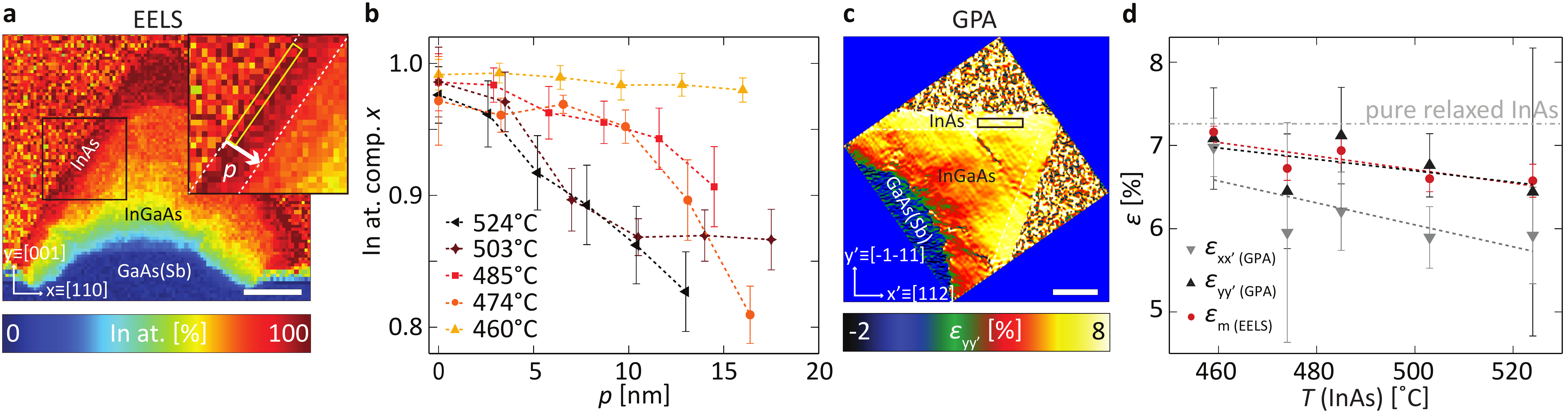}
\caption{Influence of the growth temperature on the InAs channel composition. (a) In atomic composition EELS map (relative to Ga, scale bar 50 nm) of the sample with InAs channel grown at~485~$^{\circ}$C. Inset: zoomed area of the channel indicating the procedure to obtain data points in (b). (b) In atomic composition~\emph{x} extracted at position~\emph{p} across the InAs channel grown at five different temperatures. For each growth temperature six data points are extracted. Each data point is taken at the position~\emph{p} and averaged over~100~nm~(see inset in (a)). The averaging gives the standard deviation as error bars. (c) An example of the out-of-plane lattice mismatch map ($\varepsilon_\mathrm{yy'}=\Delta d_{(\bar{1}\bar{1}1)}/d_{(\bar{1}\bar{1}1)}$) obtained with GPA (scale bar~50~nm, GaAs is the reference) for the nanowire from (a). x-axis is rotated~54.75$^{\circ}$ with respect to the original direction: $\mathrm{x'}=\mathrm{x}+54.75^{\circ}$ ($\mathrm{x'}$ is parallel to the InAs/InGaAs interface, and thus to the [112] crystallographic orientation, $\mathrm{y'}$ is parallel to the [$\bar{1}\bar{1}$1] crystallographic orientation). (d)  $\varepsilon_\mathrm{yy'}$ and $\varepsilon_\mathrm{xx'}$ as a function of the InAs growth temperature extracted from GPA. Each data point is averaged over a 50$\times$10 nm$^{2}$ box and presented with standard deviations as error bars. We also plot $\varepsilon_{m}$ calculated from the EELS composition assuming Vegard's law. In both cases, $\varepsilon$ increases with decreasing InAs growth temperature. The dashed lines are a liner fit to the experimental data and are guides to the eye.}
\label{Fig4}
\end{figure*}

Similar phenomena of diffusion and material intermixing activated at high growth temperatures in strained epitaxial systems have earlier been reported in quantum dots~\cite{Joyce1998Dec, Liao1999Dec, Chaparro2000Jun}, free-standing axial nanowires, and in-plane nanowires grown on nanomembranes~\cite{Andrade2012Nov,Beznasyuk2017Aug, Friedl2018Apr, Friedl2020May}. Among possible mechanisms, bulk diffusion and surface diffusion have been suggested. In the case of Ge/Si quantum dots, it has been shown by both experiment and simulations that at standard growth temperatures the material for intermixing is provided by surface erosion of the Si substrate creating depressions around the growing system~\cite{Liao1999Dec,Chaparro2000Jun,Tu2007Feb}. As the bulk diffusion of Ga is negligible at the growth temperatures used in the current work~\cite{Mussler2005Aug}, we conclude that Ga is supplied from the surface of the GaAs(Sb) buffer and/or the underlying GaAs substrate, consistent with the observed surface erosion, similar to the Si/Ge case.

Overall, from EELS analysis we find that the sample grown at the highest growth temperature has on average the lowest In concentration in the InGaAs buffer for both [1$\bar{1}$0]- and [100]-oriented nanowires, indicating that the strain-driven Ga-In material intermixing is promoted by elevated temperatures. 

To investigate the crystal quality of the samples, we employ Geometric Phase Analysis (GPA) on high-resolution HAADF-STEM images. Fig.~\ref{Fig3}d,e shows GPA rotational maps (planes bending with respect to the substrate). An array of misfit dislocations is seen at the InGaAs/GaAs(Sb) interface. We also find~1-2 stacking faults per nanowire cross-section. They originate at the dislocated interface on the sides of the nanowire. Some part of the mismatch strain is released elastically via a~$\sim$2$^{\circ}$ bending of planes close to the nanowire corners. The latter is possible owing to free side walls of SAG nanowires similar to free-standing nanowires~\cite{Beznasyuk2020Jul}.  
On the other hand, the InAs/InGaAs interface, highlighted with dashed lines exhibits a high crystalline quality. Over four analyzed samples, we find only~1-2 misfit dislocations over the entire InAs/InGaAs interface. These are always found at the corners of the nanowires, between the Ga-rich flames and the InAs channel. 

Our findings highlight the importance of the In-rich InGaAs buffer layer introduced between the InAs channel and the GaAs substrate. With the composition range obtained in this work, we estimate that the lattice mismatch between the InAs channel (containing on average~10\% Ga as extracted with EELS) and the In$_x$Ga$_{1-x}$As buffer reduces from~1\% to~0.47\% by reducing the buffer growth temperature. The latter allows to extend the critical thickness from only a few monolayers for InAs/GaAs interfaces to~$\sim$18~nm for InAs/InGaAs interfaces before misfit dislocations are introduced (for pure edge dislocations \cite{Fitzgerald1991Nov}). This is an important step forward toward high-quality InAs SAG nanowires \cite{Krizek2018Sep,Aseev2019Jan}.

To corroborate on the composition differences between the samples, we use XRD. In contrast to EELS, where only local nanowire composition is measured through a transversal cut of 50-100 nm thickness, XRD allows to access an average In composition~\emph{x} of an entire layer in an array of nanowires (Supplemental Material~S4). Figure~\ref{Fig3}f shows~\emph{x} in the buffer as a function of the InGaAs buffer growth temperature for the [1$\bar{1}$0]-oriented nanowires. The maximum value of~\emph{x}=0.84 is reached at~520~$^{\circ}$C and it gradually decreases to~\emph{x}=0.76 at~541~$^{\circ}$C. A similar trend is observed for the [100]-oriented nanowires. We also plot full-width at half-maximum (FWHM) from XRD peaks. The broadening of the peaks can be caused by several reasons including the strain and the distribution in the chemical composition. In our experiment the InGaAs buffers are mainly plastically relaxed and the changes in the FWHM reflect the distribution in the chemical composition within the buffer (compositional variations visible in Fig~\ref{Fig3}b and Supplemental Material~S4). There is a clear tendency toward increased FWHM of~\emph{x} at higher growth temperatures. This shows that lower growth temperatures are beneficial for both higher In concentrations and compositional uniformity of the InGaAs buffer.

We then used XRD to extract an In composition from the InAs channels. We find~\emph{x}$\sim$0.89 for all the samples in a good agreement with EELS results.

\begin{figure} [t!]
\includegraphics[width=1\linewidth]{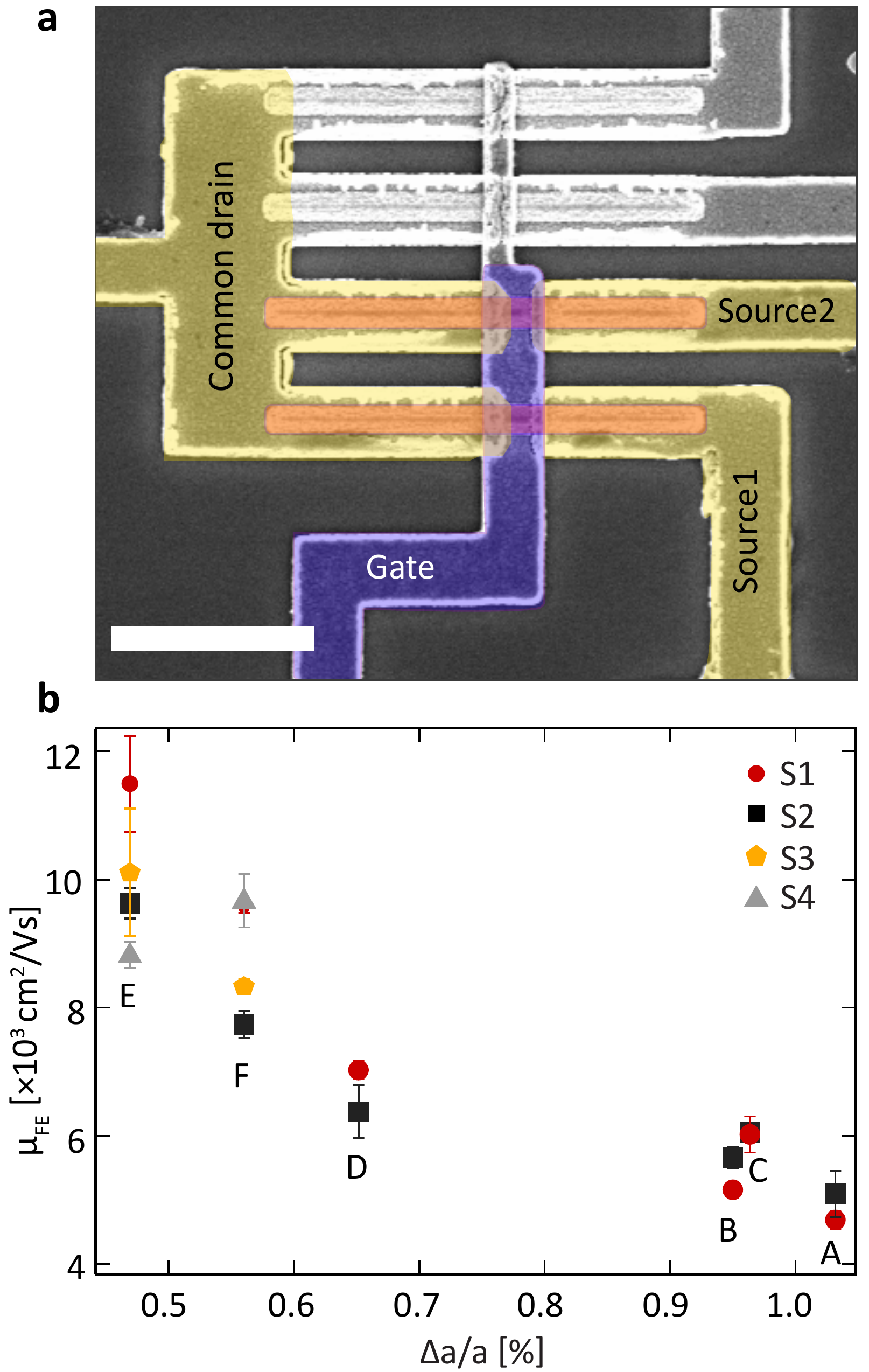}
\caption{The role of the InAs/InGaAs interface quality on the electrical properties of InAs/InGaAs/GaAs(Sb)
nanowires. (a) False-colored SEM image of field-effect devices (2 $\mu$m scale bar). Measured nanowires appear in
red, Au/Ti contacts - in yellow, the top Au/Ti gate - in violet. (b) Electron mobility, $\mu_{FE}$, as a function of the InAs/InGaAs lattice mismatch, $\Delta$a/a, for six samples collected in Table~\ref{tbl:1}: A-F. Two nanowires per sample for samples A-D are measured: S1 and S2 both having~500~nm channel length (corresponding to Source1 and Source2 in (a)) and four nanowires per sample for samples E,F are measured: S1, S2, S3, S4, all having~500~nm channel length.}
\label{Fig5}
\end{figure}


\subsection{InAs growth optimization}

\begin{table*}[!]
\centering
\caption{List of growth parameters and transport data. $W$, $V_{th}$, $\mu$, $n_{2D}$, $S_1$ and $S_2$ denote nanowire width, threshold voltage, maximum electron mobility, carrier concentration, outer and inner nanowire, respectively. $n$ is estimated at zero gate voltage $V_g$=0 via the formula $n_{2D}=C\Delta V/Ae$~\cite{Radisavljevic2013Sep}, where $C$ is the capacitance found with the finite element simulations as described in~\cite{Krizek2018Sep}, $e$ is the elementary charge, $A=LW$ is the surface area, and $\Delta V$=$V_g$-$V_{th}$. Data for two more nanowires from samples E,F can be found in Supplemental Material~S9.}
\label{tbl:1}
\begin{tabular}{|c|c|c|c|c|c|c|c|c|c|c|}
\hline
\multirow{2}{*}{Sample} &
\multirow{2}{*}{$T_{InGaAs}$, $^{\circ}$C} & \multirow{2}{*}{$T_{InAs}$, $^{\circ}$C} & \multicolumn{2}{c|}{$W$, nm} & \multicolumn{2}{c|}{$V_{th}$, V} & \multicolumn{2}{c|}{$\mu$, cm$^2$V$^{-1}$s$^{-1}$} & \multicolumn{2}{c|}{$n_{2D}$, 10$^{12}$ cm$^{-2}$} \\ \cline{4-11} 
                           &&& S$_{1}$  & S$_{2} $  &S$_{1}$ & S$_{2}$ & S$_{1}$ & S$_{2}$ & S$_{1}$ & S$_{2}$\\ 
                           \hline
                           
A & 539 & 520 & 270 & 270   & -1.61 &-0.84 & 4859 & 5541 &   7.76 & 4.05 \\ \hline
B & 535 & 520  & 280 & 290 & -1.23 & -1.12 & 5271 & 5881 & 5.69 & 5.02\\ 
\hline
C & 533 & 520  & 270 & 270 &-1.09 &	-1.49 & 6283 & 6165 & 5.23 & 7.42 \\ 
\hline
D & 529  & 520 & 250 & 250 &-1.72 &	-1.11 & 7095 & 6360 & 8.93 & 5.74 \\ \hline
E & 520      & 520       & 290       & 290   &-2.00	&-2.00 & 12550  & 9949  & 8.95 & 8.95\\\hline 
F & 520  & 474  & 250 & 250 & -1.83	 & -1.65 & 8412 & 10064 & 9.48 & 8.57\\ \hline
\end{tabular}
\end{table*}

As we saw before, the Ga-In material intermixing takes place not only in the bulk of the InGaAs buffer layer but also in the InAs channel (Fig.~\ref{Fig3}b,c). To improve the composition homogeneity of the channel, we now optimize the growth temperature of the InAs layer. For this, five samples are grown in the temperature range between~460-524~$^{\circ}$C (see SEM images in Supplemental Material~S5). The InGaAs growth temperature is fixed at~520~$^{\circ}$C for all samples as it is found from the previous section to be the optimum.  

We first begin by analysing the compositional differences between the InAs nanowire channels by using EELS (Fig.~\ref{Fig4}a and Supplementary Material~S5). The In composition~\emph{x} (with respect to Ga) is extracted as a function of position~\emph{p} across the channel as shown in Fig.~\ref{Fig4}b (see inset to Fig.~\ref{Fig4}a for definition of p). We note that for all samples the outermost layer (\emph{p}=0) is pure InAs owing to surface segregation~\cite{Moison1989Jan, Tu2007Feb}. However, the composition of the inner layer is significantly different. High growth temperatures result in highly nonuniform compositions with~\emph{x} decreasing down to $\sim$83\% across the channel. The relatively large error bars (up to~3\%) are a consequence of composition broadening along the channel as well. On the contrary, when the temperature is as low as~460~$^{\circ}$C, the In composition remains above~97\% across the entire channel and error bars decrease to~$\sim$1\% demonstrating high composition homogeneity along the channel.

Based on the values of~\emph{x} extracted with EELS we obtain the lattice mismatch between the InAs channel and the GaAs buffer: $\varepsilon_m=(a_{In_xGa_{1-x}As}-a_{GaAs})/a_{GaAs}$ extrapolating the In$_x$Ga$_{1-x}$As lattice parameter~$a_{In_xGa_{1-x}As}$ assuming Vegard's law (Fig.~\ref{Fig4}d)~\cite{Adachi2006}. $\varepsilon_m$ reaches its maximum value of~7.14$\pm$0.07\% at~460~$^{\circ}$C corresponding to the pure InAs material (\emph{x}=1). To validate our measurements, we extract the
lattice mismatch between InAs and the GaAs(Sb) buffer layer using GPA applied on atomic resolution HAADF STEM images. The in-plane ($\varepsilon_\mathrm{xx'}$) and out-of-plane ($\varepsilon_\mathrm{yy'}$) lattice deformations are plotted in Fig.~\ref{Fig4}d. An identical trend can be seen: $\varepsilon_\mathrm{xx'}$ and $\varepsilon_\mathrm{yy'}$ increase with the decrease in growth temperature. Note, that $\varepsilon_\mathrm{xx'}$ is smaller than $\varepsilon_\mathrm{yy'}$ suggesting that the InAs layer remains compressively strained in the plane of the interface. The latter causes the expansion of the out-of-plane lattice constant owing to the Poisson effect~\cite{Pohl2013}.

We then analyzed the crystal quality of InAs/InGaAs interfaces with GPA. Overall,~1-2 misfit dislocation(s) per nanowire at the InAs/InGaAs interface along the [1$\bar{1}$0] direction are observed, except for the sample grown at~474~$^{\circ}$C. It shows a more defective interface with~4 misfit dislocation(s). By decreasing the InAs growth temperature, the Ga content in the channel is reduced increasing the InAs/InGaAs lattice mismatch. However, based on only one cross-section per sample it is impossible to draw conclusions whether nanowires from the lowest temperature samples are more defective as compared to the high temperature samples. 

\subsection{Electrical transport properties}
Having shown that the In content in the InGaAs buffer and the InAs active region increases by decreasing the growth temperature, we examine if it also affects the electrical properties. Devices are fabricated directly on the growth substrates. Prior to depositing Ti~5~nm/Au~250~nm ohmic contacts, we remove the native oxide from the nanowires by Argon ion milling at 15 W for~150~s. The time of the milling is carefully adjusted to make sure that the InAs is not etched away. We then deposit~10~nm of HfO$_2$ at~90~$^{\circ}$C by means of atomic layer deposition to separate the top gate from the contacts. Finally, Ti~5 nm/Au~250~nm top gates are deposited. The patterning of contacts and top gates is done with electron beam lithography. Fabricated devices are then cooled down in a cryogen-free DynaCool physical property measurement system (PPMS) with a base temperature of~1.7~K.

We measure single [1$\bar{1}$0]-oriented InAs/InGaAs/GaAs(Sb) nanowires (Figure~\ref{Fig5}a) with~500~nm channel length, $L$, and~250-290~nm channel width, $W$, depending on the sample (Table~\ref{tbl:1}). The field-effect mobility, $\mu_{FE}$, is extracted by fitting measured conductance, $G$, versus applied top gate voltage, $V_g$, as described in Ref.\cite{Gul2015May} (Supplemental Material~S6). Figure~\ref{Fig5}b shows $\mu_{FE}$ as a function of the InAs/InGaAs lattice mismatch, $\Delta$a/a, calculated assuming Vegard's law and using~\emph{x} extracted with XRD for both channel and buffer. For the nanowires grown with a different buffer growth temperature (samples A-E, Table~\ref{tbl:1}), $\mu_{FE}$ increases from~4700$\pm$100~cm$^2$V$^{-1}$s$^{-1}$ for~$\Delta$a/a=1.03\% (T$_{InGaAs}$=540$^{\circ}$C, sample A) to~11500$\pm$700~cm$^2$V$^{-1}$s$^{-1}$ for~$\Delta$a/a=0.47\% (T$_{InGaAs}$=520$^{\circ}$C, sample E). The carrier concentration,~$n_{2D}$, ranges from~4.05 to~9.5~$\times$10$^{12}$~cm$^{-2}$ as estimated at~$V_g$=0.

From previous reports considering 15-20 nm wide InAs quantum wells at high carrier densities, it is known that there are primarily two scattering mechanisms affecting the mobility at low temperature: the interface roughness and the alloy disorder\cite{Shojaei2016Dec}. Samples A-E have the same composition of the InAs channel as accessed with EELS and XRD resulting in the same order of alloy disorder in the channel. We therefore attribute the mobility dependence in Fig.~\ref{Fig5}b to an effect of surface roughness and our results suggest that the quality of the InAs/InGaAs interface is significantly improved by growing InGaAs buffer layers at low growth temperatures. Indeed, the lattice mismatch reduces from~1\% to~0.47\% for the high-temperature and low-temperature sample, respectively. Given that the critical thickness below which dislocations-free interfaces are achieved for~1\% lattice mismatch is only~$\sim$6~nm, the 15-20 nm InAs layer along the nanowire is expected to be defective for the samples grown with high buffer temperature. On the contrary, for the low buffer temperature sample with~0.47\% lattice mismatch the critical thickness is~$\sim$18~nm and interfaces without dislocations are expected. The interface improvement directly translates to improvement in the electrical properties of the InAs layer in agreement with previously reported results for InAs/AlGaSb heterostructures\cite{Shojaei2015Jun}.

An alternative interpretation of the trend observed in Fig.~\ref{Fig5}b could be that a proportion of the conduction takes place in the InGaAs layer, where the reduction of alloy disorder and increased In concentrations at lower temperatures could potentially explain the observed mobility dependence. However, this situation is highly unlikely, since electron transport in InAs occurs in a surface accumulation layer with a depth on the order of 15-20 nm~\cite{King2010Jun,Schuwalow2021Feb}, less or equal to the InAs channel thickness. To confirm this, we performed simulations of the band structure using a 2D Schr\"{o}dinger-Poisson model (Supplemental Material~S7). The FWHM of the wavefunction is 11 nm at zero gate voltage suggesting that the bulk of conduction is confined to the InAs layer, with only a small part of the wavefunction tail overlapping into the InGaAs buffer. To support this conclusion, we measured devices \emph{without} an InAs layer (Supplemental Material~S8). We found that inducing carriers in InGaAs/GaAs(Sb) devices required higher $V_g$ than for InAs/InGaAs/GaAs(Sb) devices, consistent with the notion that at the gate voltages used to calculate mobility in Fig.~\ref{Fig5}b electrons were not occupying the InGaAs. In all, the mobility gains are attributable to reduced InAs/InGaAs interface roughness and reduced InAs dislocation density.

Finally, to check if improvement in the composition purity of the InAs channel affects the mobility, we measure sample F from the InAs sample series (T$_{InAs}$=474$^{\circ}$C) and compare it with sample E (T$_{InAs}$=520$^{\circ}$C). The sample grown at the lowest temperature (T$_{InAs}$=460~$^{\circ}$C) was not chosen for transport measurements because of a high density of parasitic clusters on the oxide mask. These are often merged with nanowires affecting their morphology. Having measured 4 nanowires per sample, we find that $\mu_{FE}$ remains in the range ~8000-10000~cm$^2$V$^{-1}$s$^{-1}$ (Fig.~\ref{Fig5}b) comparable to the high temperature sample E. These results support the conclusion that $\mu_{FE}$ is mainly limited by the InAs/InGaAs interface quality. Sample F has on average higher In content in the channel than sample E resulting in slightly increased lattice mismatch with the underlying buffer layer. Our data favorably agree with existing literature on InGaAs/GaAs bulk materials\cite{Chin1991Oct}. 

\section{Conclusion}

In conclusion, we have successfully optimized InAs/InGaAs SAG nanowires on GaAs(001) substrates doubling their electron mobility. The carrier mobility obtained in this work is higher than state-of-the-art values for SAG of InAs/GaAs and In$_{0.5}$Ga$_{0.5}$As/GaAs  nanowires and nanofins~\cite{Krizek2018Sep,Friedl2018Apr,Friedl2020May,Seidl2019Jul} demonstrating an advancement toward realizing high quality gatable InAs quantum channels based on the SAG approach. We improved the nanowire/substrate interface quality by substituting conventional thermal annealing for atomic hydrogen for the native oxide removal and by introducing a metamorphic In$_{x}$Ga$_{1-x}$As buffer layer between the InAs channel and the GaAs substrate. The Ga-In material intermixing was inhibited by reducing the growth temperature of InGaAs and InAs. We observed that In-rich InGaAs buffer layers grown at reduced temperatures result in improved InAs/InGaAs interfaces owing to the reduced lattice mismatch. The latter is found to be a crucial factor for enhancing the electron mobility of InAs/InGaAs SAG nanowires.

\medskip 
\textbf{Data availability.}
All data needed to evaluate the conclusions in the paper are present in the paper and in the Supporting Information.

\begin{acknowledgments}
The project was supported by Microsoft Quantum, the European Research Council (ERC) under Grant No. 716655 (HEMs-DAM), the European Union Horizon 2020 research and innovation program under the Marie Skłodowska-Curie Grant No. 722176. The authors acknowledge Dr. Keita Ohtani for technical support and fruitful discussions. D.V.B. is grateful to Dr. Juan-Carlos Estrada Salda\~{n}a for careful reading of the manuscript. The authors thank Francesco Montalenti, Marco Albani and Leo Miglio for scientific discussions. ICN2 acknowledges funding from Generalitat de Catalunya 2017 SGR 327. ICN2 is supported by the Severo Ochoa program from Spanish MINECO (Grant No. SEV-2017-0706) and is funded by the CERCA Programme/Generalitat de Catalunya. Part of the present work has been performed in the framework of Universitat  Autònoma de Barcelona Materials Science PhD program. The HAADF-STEM microscopy was conducted in the Laboratorio de Microscopias Avanzadas at Instituto de  Nanociencia de Aragon-Universidad de Zaragoza. M.C.S. has received funding from the  European Union’s Horizon 2020 research and innovation programme under the Marie Sklodowska-Curie grant agreement No 754510 (PROBIST). We acknowledge support from CSIC Research Platform on Quantum Technologies PTI-001.
\end{acknowledgments}

\bibliography{Bibliography}
\bibliographystyle{apsrev.bst}

\end{document}